\documentclass[onecolumn,aps,pra,10pt]{revtex4-2}
\usepackage[utf8]{inputenc}
\usepackage{graphicx}
\graphicspath{{./}}
\usepackage{lmodern}
\usepackage[T1]{fontenc}
\usepackage{geometry}
\usepackage{csquotes}
\usepackage{siunitx}
\DeclareUnicodeCharacter{00B0}{°}
\DeclareUnicodeCharacter{04AB}{c}

\begin{document}
\title{Ultra-diluted gas transmittance revisited}
\author{Jakub M. Ratajczak}
\affiliation{Centre of New Technologies, University of Warsaw, Poland\\(www.smearedgas.org)}

\date{July 2021, rev. September 2022}

\begin{abstract}
The paper analyzes a model of optical transmittance of ultra-diluted gas, considering gas particles' non-locality and the quantum effect of their wave function spreading derived from solving the Schrödinger equation for a free particle. The analysis does not depend on a particular form of the wave function, but it assumes the reality of wave function. Among others, we show conserved mass gas clouds may become significantly more transparent than predicted by classic transmittance laws. This unexpected phenomenon is possible because mass conservation is governed by the sum of probabilities, while the Markov chain's product of probabilities controls the transmittance. Furthermore, we analytically derive the upper limit the closed system transmittance may grow and demonstrate a boundless, open gas cloud transmittance may grow up to 100\%. Finally, we show the impact on interpretations of quantum mechanics. The model is naturally applicable in deep space conditions, where the environment is sparse. Furthermore, the model responds to dark matter requirements.
\end{abstract}
\maketitle

\section{Introduction}
The Beer-Lambert exponential transmission law \cite{Bouguer1729} \cite{A.D.McNaught1997} describing attenuation of monochromatic light by the homogeneous, the not very dense medium is well known for almost three centuries. Despite developing newer, more advanced transmittance models, it still applies to quantitative spectroscopy \cite{Bernath2016}, rarefied gases, and astrophysical measurements. All these models rely on an assumption of attenuating particle locality. However, an increasing number of experiments \cite{Handsteiner2017} \cite{Rauch2018} convince us that the underlying theory of Quantum mechanics is not a local realistic theory \cite{Einstein1935} \cite{Musser2016}. There is one more assumption in most ''classic`` transmittance models: a light detector is a macroscopic apparatus. Quantum mechanics is one of the most fundamental theories, so it is necessary to check whether these two assumptions limit the scope of applicability of classic transmittance models. 

Quantum spreading is an effect that involves spontaneous spatial smearing of the $\Psi$ wave function over time. It leads to the spreading of the $|\Psi|^2$ probability density of any reaction of a physical object described by such a function. It comes directly from the free particle Schrödinger equation solution \cite{Shankar2011}. Assuming wave function reality \cite{Hobson2013} \cite{Long2018} we apply this solution to each gas particle independently during its free time between successive collisions. We proposed a kind of ''smeared gas`` model. It leads, together with the assumption of non-locality, to a new model   of electromagnetic transmittance of thin gases. One of the predictions of this model is that the measured optical transmittance depends, among others, on the size of the detector used and the duration of the particle mean free time. The classic ''local`` approach to transmittance, e.g., the Beer-Lambert law, does not predict such dependencies.

This paper presents a deeper analysis of the smeared gas transmittance model \cite{Ratajczak2020b}. We analyze open and closed systems. We show that transmittance may rise, thanks to spontaneous particle spread, even in the closed system but only up to some limit. We analytically derive this limit. We show that displacement of the measurement axis in relation to the cloud mass center may affect transmittance measurement. The $G$ parameter of the smeared gas transmittance model is analyzed more thoroughly. In the end, we briefly address the possibility of distinguishing quantum mechanics interpretations using the outcome of the model.

\section{Assumptions}
There are just a few assumptions for the model. Gas particles are independent of each other, and they are non-local wave functions (not necessary) of the same type. Gas is not relativistic, so the Schrödinger equation applies. Particle distribution is homogenous, and wave functions differ by position only. The light detector is of some finite size. The paper \cite{Ratajczak2020b} describes those assumptions in detail.

\begin{figure*}
\includegraphics[width=\textwidth]{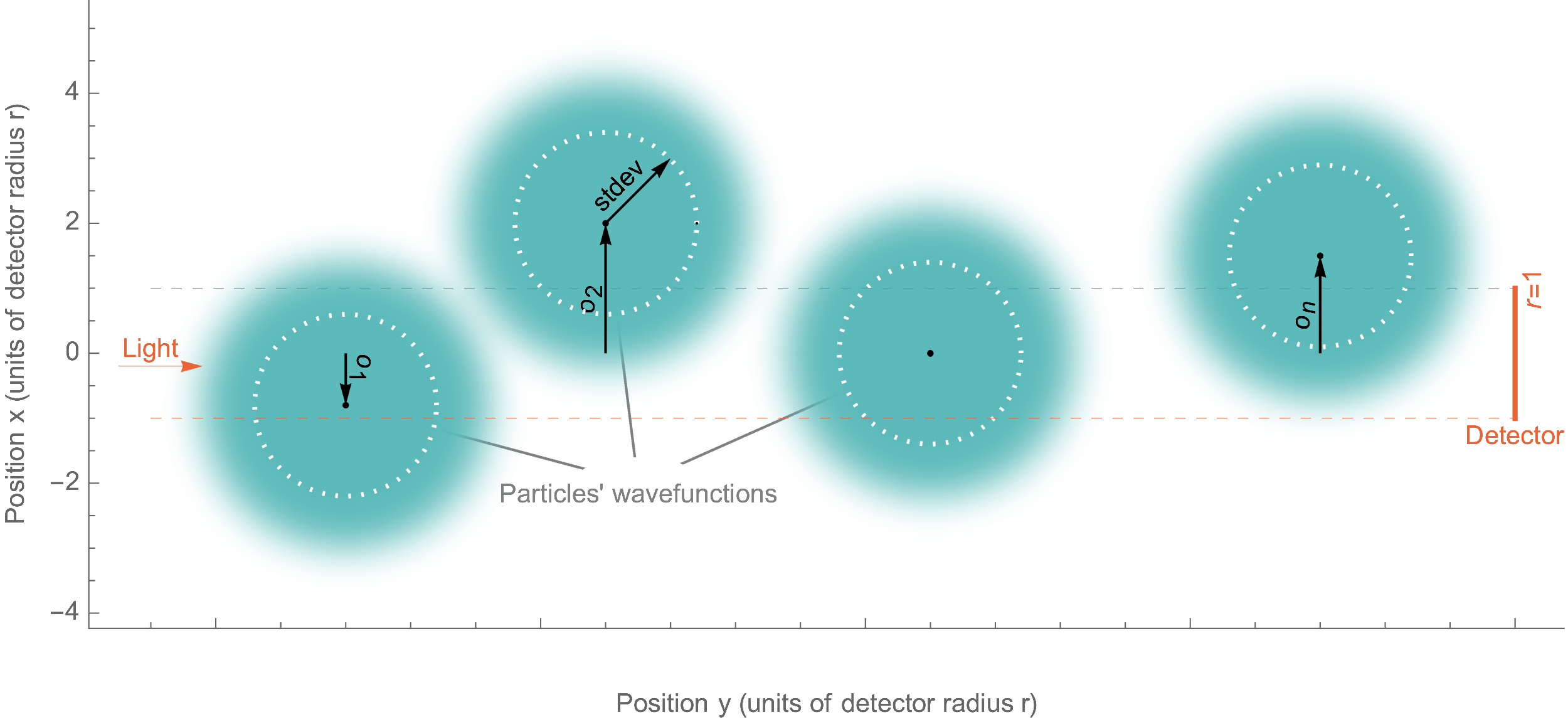}
\caption{A sample diagram shows a gas cloud made of a few 2D particles. Their wavefunctions have the Gaussian distribution with uniform standard deviation $stdev$. Each particle is offset from the $x=0$ axis by $o_n$. The light direction, position of the light detector, and visibility tunnel (volume of integration) are marked red. The detector is centered and placed parallel to the $X$ axis. Units are selected to have the detector diameter $r=1$.} \label{fig:particlesChart}
\end{figure*}

In this paper, first we examine a gas cloud made of two-dimensional particles, see Fig.~(\ref{fig:particlesChart}). We show later how the two-dimensional case upgrades to three dimensions, not affecting conclusions drawn. 

Most of the analysis considers a detector of some constant diameter equal to $2r$. The detector radius $r$ is used as a length unit $r=1$ across the paper. In real-world cases, r may be from microns to meters. We assume 100\% detector efficiency without loss of generality. 

We assume a simple measurement setup. Monochromatic light propagates perpendicularly to the detector from a source of precisely identical shape and size as the detector. The volume between them is called a ''visibility tunnel``. This tunnel is the only area where photon absorption may affect the number of photons (not) counted by the detector. We assume the detector is big enough (macroscopic), so the visibility tunnel doesn't get wider due to non-classic photon trajectories. Both the source and the detector are far from the cloud. See the ''astronomical setup`` considerations in \cite{Ratajczak2020b}.

We interpret each individual gas ''particle`` wave function realistically: ''$\Psi(x)$ is a spatially extended field representing the probability amplitude to interact at $x$ rather than an amplitude for finding, upon measurement, a particle.`` \cite{Hobson2013}. We don't constrain the exact form of the wave function. The normal distribution is used later in the text. It fulfills the above requirement, and it is the free particle Schrödinger equation solution. Also, it provides a convenient spread measure, namely the standard deviation ($stdev$). 

For a free particle distribution derived from the Schrödinger equation, the standard deviation depends on particle free time (e.g. (4) in \cite{Ratajczak2020b}). We assume low cloud density, thus a non-decohering environment to let particles evolve free for some time, so wave functions reach substantial spread spontaneously. For simplicity, the same spread for all particles in the cloud is considered: the same standard deviation for all probability distributions. However, combining multiple transmittance equations may relax the requirement for uniform distributions if necessary.

Also, the presented model neither depends on the idea of wave function collapse nor applies directly to the quantum measurement problem. We are not analyzing what happens to a wave function after absorption. 

For text simplicity, we describe as ''absorption`` all types of events that may happen to a photon on its way to a detector, namely either scattering or absorption. 

A single particle cross-section ($\sigma$) must be smaller than the detector area $\sigma \ll r^2$, which is the case for any atomic or molecular gas and common macroscopic detectors.

\subsection{The $G$ coefficient revisited}

This paper discusses a gas cloud optical transmittance model dependent on the size of a detector and gas particles' wave function spread. However, to relate its predictions to actual physical environments, we also need to include properties such as the density of a gas cloud, cloud thickness, and (a single particle) attenuation cross-section. An appropriate inclusion will adjust the transmittance equation to a given setup, allowing quantitative, experimental predictions.  

In the proposed model, the $G$ coefficient plays the role of a normalizing factor \cite{Ratajczak2020b}. Its value depends on the physical properties of the scattering medium: wavelength-dependent particle cross-section, cloud thickness, and cloud density. To apply the $G$ factor to the probabilistic model discussed in the paper, it must encode those properties as follows. $G$ indicates how much ($0<G\leq1$) of the detector's surface is ''covered`` by the obscuring cloud in the classical limit, i.e., gas particles wavefunctions are highly localized, and the detector is macroscopic: $r \gg stdev$. Such a definition is required to guarantee identical quantitative results of the classical model \cite{Bouguer1729} and our (in the classical limit, where $r \gg stdev$). In other words, $G$ is a probability that a photon (that would reach the detector in the absence of the cloud) is not absorbed/scattered somewhere in the cloud. And this is just a probabilistic complement of the classic transmittance $TR_{cl}$ of a cloud: 
\begin{equation}
G=1-TR_{cl}~.
\label{GDefinition}
\end{equation}

Let's recall \cite{A.D.McNaught1997} $TR_{cl}=e^{-nl\sigma}$, where $n$ is the particles number density, $l$ is the cloud thickness in the measurement direction (light length) and $\sigma$ is particle attenuation cross-section. There are other popular ways of quantifying opacity, namely optical depth ($\tau=nl\sigma$) or absorbance ($ABS$). They are related to each other: $TR_{cl}=e^{-\tau}=10^{-ABS}$ so we can express $G$ in terms of them:
\begin{equation}
G=1-e^{-\tau}=1-10^{-ABS}
\end{equation}
It is clear the $G$ coefficient is limited: $0<G\leq1$ as required. We see the $G$ coefficient is yet another way of expressing a classical optical depth of a gas cloud. Moreover, it is directly related to optical depth or absorbance by a simple arithmetic relation.

We showed what is $G$ for a homogenous cloud and monochromatic light. It is a simplification but still useful for many applications, e.g., spectroscopy and astrophysics. If necessary, one may extend the model presented here to heterogeneous clouds and many wavelengths the same way classic homogenous and monochromatic absorbance extends to more complicated cases.

In the following examples and figures, we put $G=0.7$. It is a case of a cloud with 30\% transmittance $TR_{cl}=1-0.7$ or optical depth $\tau=-ln(TR_{cl})\approx1.20$ or absorbance $ABS=-log_{10}(TR_{cl})\approx0.52$. We chose this particular value because it corresponds to the typical transmittance in the conducted experiment \cite{Ratajczak2021}.

\section{A single particle transmittance}
This section shows how a single particle spread affects the absorption detection rate depending on the detector size and position. It is a kind of the simplest, single-particle gas cloud. We introduce probability distribution, detection rate and illustrate.

We will be interested in finding a particle in a given volume of detectability tunnel perpendicular to the detector plane. To simplify calculations, we can project a 2D particle onto the detector plane thanks to distribution symmetry. This way, we get 1D normal distribution $P$:
\begin{equation}
P(x)=|\Psi(x)|^2=\frac{1}{ \sqrt{2\pi} stdev }exp \left( \frac{-x^2}{2stdev^2} \right)
~, \label{ProbabilityDistribution}
\end{equation}
where $stdev$ is particle standard deviation.

The probability $P_v(o)$ of finding a particle in a given volume of detectability tunnel $(o-r)<x<(o+r)$ is the integral:
\begin{equation}
P_v(o)=\int_{o-r}^{o+r}P(x)dx=\frac12\left[ erf \left( \frac{o-r}{\sqrt2 stdev} \right) - erf \left( \frac{o+r}{\sqrt2 stdev} \right) \right]
~, \label{FindProbability}
\end{equation}
where $erf()$ denotes the Gauss error function, and $o$ is the distance (offset) from the visibility tunnel axis to the particle. However, this probability is not equivalent to the probability of absorbing a photon by this particle in this volume. Absorption probability depends additionally on the physical properties of a cloud, as discussed in the previous section: particles' cross-section, density, and thickness. This is what the $G$ coefficient is responsible for. It encodes the probability of passing a photon from the source to the detector - in the presence of a classical cloud. Both events (i.e., the particle in the volume and absorbing a photon) must coincide to prevent a photon from reaching the detector. Therefore, we need to multiply both probabilities and take the complement to get the photon passing probability. This probability is, by definition, the cloud optical transmittance ($TR$):
\begin{equation}
TR(o)=1-G\,P_v(o)~. \label{TR_G}
\end{equation}

Applying  Eq.~(\ref{FindProbability}) we find the transmittance:
\begin{equation}
TR(o)=1-\frac{G}2\left[ erf \left( \frac{o-r}{\sqrt2 stdev} \right) - erf \left( \frac{o+r}{\sqrt2 stdev} \right) \right]
~, \label{TR1}
\end{equation}
as measured by the $r$-radius detector, and the particle offset from the visibility tunnel axis by $o$.

\begin{figure*}
\includegraphics[width=\textwidth]{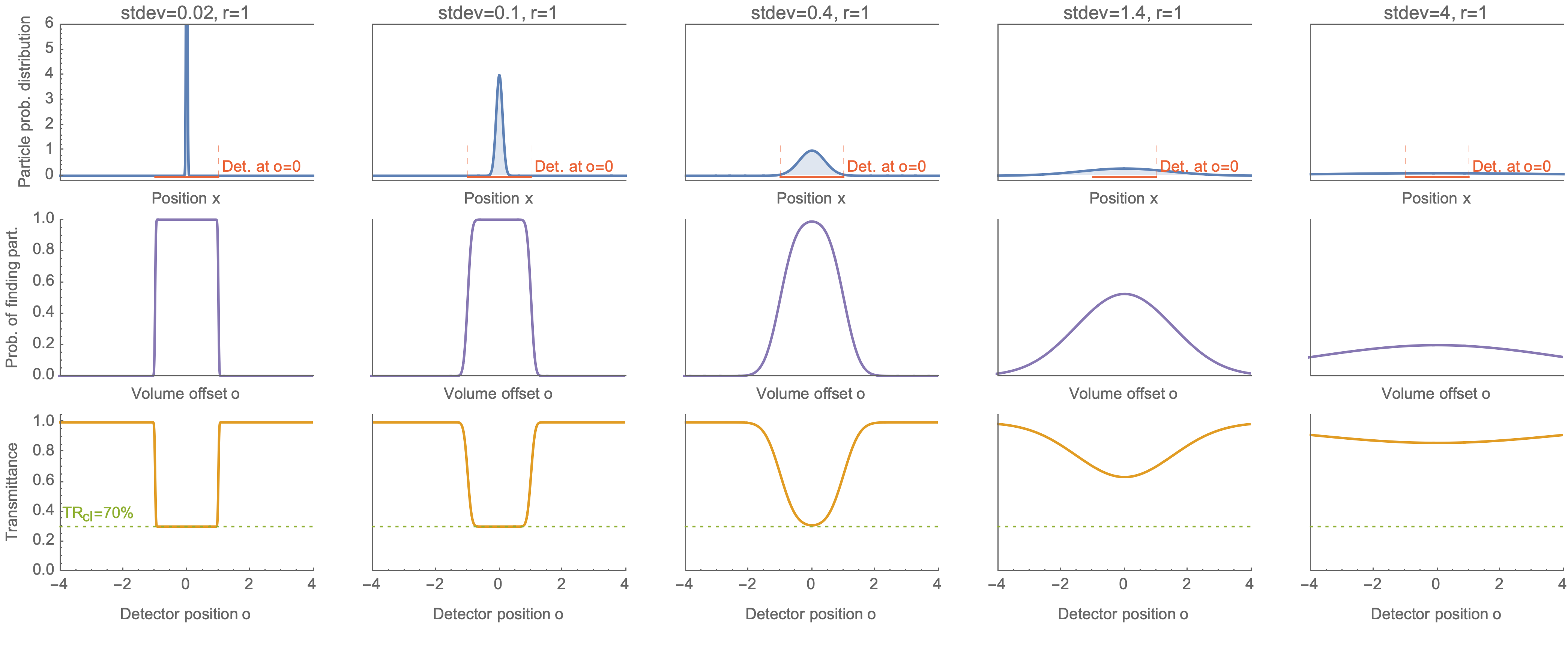}
\caption{Sample graphs for i) a single particle probability distribution, ii) probability of finding a particle within a detectability tunnel, and iii) transmittance as measured by a finite detector. The detector radius $r=1$. Each column graphs for a different $stdev$ value. i) The first row shows particle distribution $P(x)$ after Eq.~(\ref{ProbabilityDistribution}). The solid red line is a sample detector position at $o=0$. The dashed red lines show detectability tunnel (volume) boundaries. ii) The middle row shows the probability $P_v(o)$ of finding a particle within a detectability tunnel that is a volume constrained by the detector position $o$ according to Eq.~(\ref{FindProbability}). iii) The last row plots the transmittance $TR(o)=(1-G\,P_v(o))$ that would be measured by the 100\% efficient detector set at $o$, see Eq.~(\ref{TR1}) with $G=0.7$. The green dashed line shows classic transmittance $TR_{cl}$.} \label{fig:singleParticleCharts}
\end{figure*}

Fig.~(\ref{fig:singleParticleCharts}) illustrates all three equations for a couple different standard deviations. The following columns present the relationships for ever wider standard deviations of the particle distribution. The first column corresponds to a well-located particle ($stdev \ll r$), i.e., the classical situation of an ideal gas. With the increasing spread, we can see that i) the transmittance for the detector in line with the particle is always the lowest, ii) the transmittance for the detector offset further from the center increases. For an open, infinite system, the detector can be moved as far as possible. There will also be some probability of being obscured by the non-local particle.

\section{The smeared gas transmittance}
\begin{figure*}
\includegraphics[width=\textwidth]{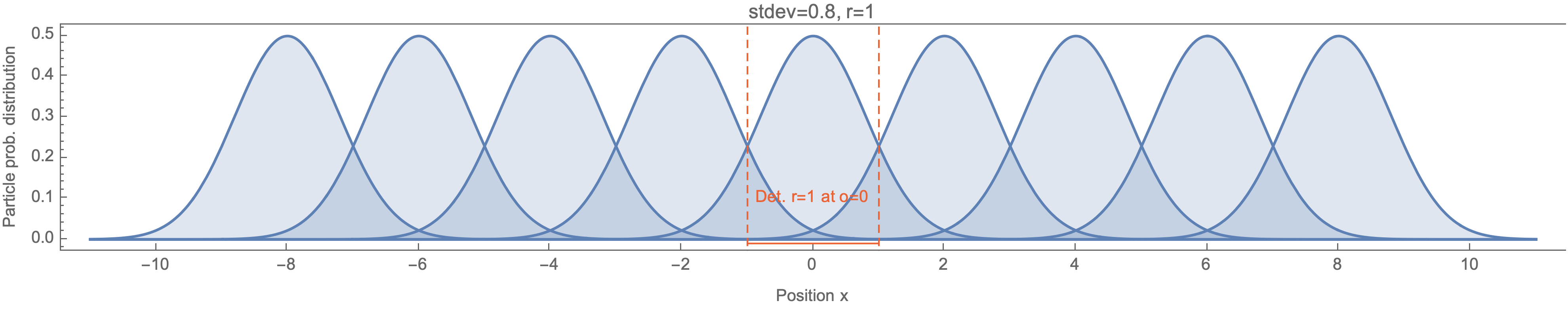}
\caption{A sample configuration of 9 identical particles distributed evenly every $2r$. The solid red line marks the detector, and the dashed red lines are visibility tunnel borders} \label{fig:manyParticlesChart}
\end{figure*}

Let us study a system of many particles: odd number of 2D particles, lined up parallel to the detector, spaced every $2r$. The detector is placed symmetrically in the middle. We will release those conditions later. An identical probability distribution gives the probability of locating each particle. Although we use the Gaussian distributions any probability distribution works, because $\int_{-\infty}^{\infty}P(x)dx=1$. This way, we do not get attached to any particular form of a wave packet. Again, to simplify calculations, we project 2D particles onto the detector plane to work with 1D distributions. Fig.~(\ref{fig:manyParticlesChart}) shows such a configuration for $N=9$ particles. The solid red line marks the detector, and the dashed red lines are the visibility tunnel boundaries.

We define dilute gas cloud transmittance $TR$ as proposed in \cite{Ratajczak2020b}. \textit{Transmittance is the probability that a photon that the detector would have detected in the absence of a cloud passes not-absorbed the entire $N$-element cloud and is detected by the detector.} Collisions with individual particles are independent, so we may consider this process as a Markov chain: 
\begin{equation}TR = \prod_{n=1}^N \big ( 1-G\,P(o_n) \big )
~, \label{TREquation}
\end{equation}
where $G\,P(o_n)$ is the probability of photon being absorbed by $n$-th gas molecule, which is offset by $o_n$ from the detector, see Eq.(5) in \cite{Ratajczak2020b}.

Now we take advantage of the periodicity. Identical chunks (of the same shape and number) of the probability distribution both leak out and flow into the visibility tunnel. Thus, we can ''unfold`` a single distribution periodically instead of considering all distributions in one place. Next, we virtually ''shift`` the detector $N$ times (by a period of $2r$) and take the product of all its positions. This way we may substitute $o_n=r(2n-N-1)/2$, and Eq.~(\ref{TREquation}) turns into:

\begin{equation}
TR = \prod_{n=1}^N \left(1-G\,P\left(r\frac{2n-N-1}{2}\right)\right)
~. \label{TRUnfold}
\end{equation}

\begin{figure*}
\includegraphics[width=\textwidth]{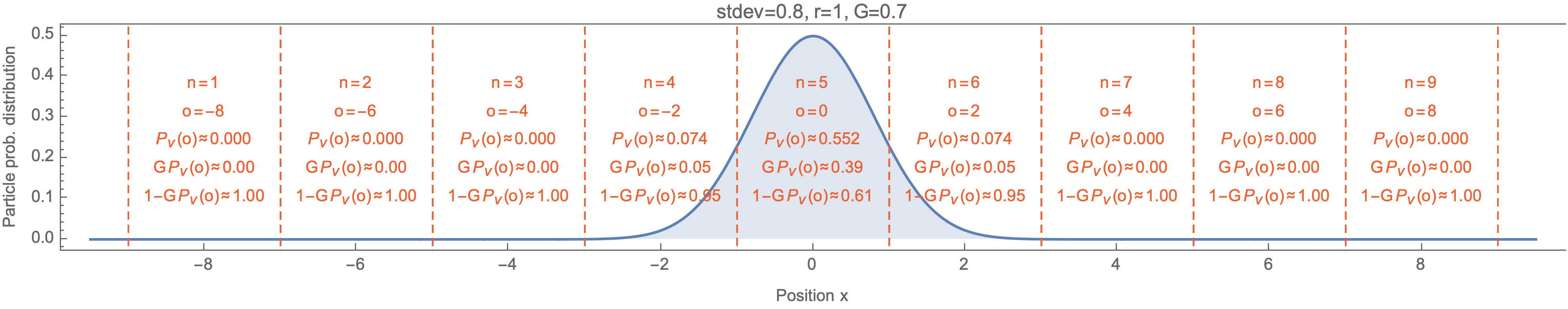}
\caption{The idea of dividing a probability distribution into many chunks, periodically every $2r$ as required by Eq.~(\ref{TRUnfold}). Values of $n$, $o$, $P_v(o)$, $G\,P_v(o)$ and $1-G\,P_v(o)$ are superimposed for every part for convenience.} \label{fig:manyDetectorsChart}
\end{figure*}

Fig.~(\ref{fig:manyDetectorsChart}) shows this idea. One distribution is divided into many chunks, periodically every $2r$. Values of $n$, $o$, $P_v(o)$, $G\,P_v(o)$ and $1-G\,P_v(o)$ are superimposed for every part for convenience.

As expected, all probabilities $P_v(o)$ sum up to 1, which means that the analyzed interval contains the entire particle. The probability is not ''leaking`` sideways. We interpret this as a conserved mass in the system. 

The following holds for $G=const$:
\begin{equation}
\sum_{n=1}^{N} P_v(o_n)=1 \Rightarrow \sum_{n=1}^{N} G\,P_v(o_n)=G=const 
~.
\end{equation}
The transmittance is the product of $1-G\,P_v(o_n)$, see Eq.~(\ref{TREquation}). The sum of its components is always constant $\sum (1-G\,P_v(o_n))=const$ as shown above. However, the constant sum doesn't guarantee the product is constant:
\begin{equation}
\sum_{n=1}^{N} a_n=\sum_{n=1}^{N} b_n \not\Rightarrow \prod_{n=1}^{N} a_n=\prod_{n=1}^{N} b_n
~.
\end{equation}

It shows that even \textit{for closed systems with mass conserved, the transmittance may change} because mass conservation depends on a sum (of some elements), but transmittance depends on a product (of the same elements). In general, the transmittance depends on how the distributions are divided. This division depends on i) the shapes of the probability distributions and ii) the width of the detector. 

The shape of the normal distribution depends only on its standard deviation. The detector size $r$ significantly influences the individual product components values for either $r\,\sim\,stdev$ or $r<stdev$, influencing the product and, finally, the measured transmittance. 

In the classical case, for well-located (ideal gas) particles and macroscopic detector, we have $stdev\,\ll\,r$. This way, all non-zero probability of a particle always gets to one chunk making all elements of the product of Eq.~(\ref{TREquation}) equal to 1, except one element. The only element less than 1 determines the value of the entire product. The product doesn't change upon changing $r$ because there will always be only one such element. So, in this case, detector size can't affect the transmittance measurement. It explains why we don't observe transmittance dependence on detector size in classic systems.

This analysis applies to any number of particles ($N$). For even $N$ the $o_n$ substitution leading to Eq.~(\ref{TRUnfold}) should be just slightly different.

\begin{figure*}
\includegraphics[width=\textwidth]{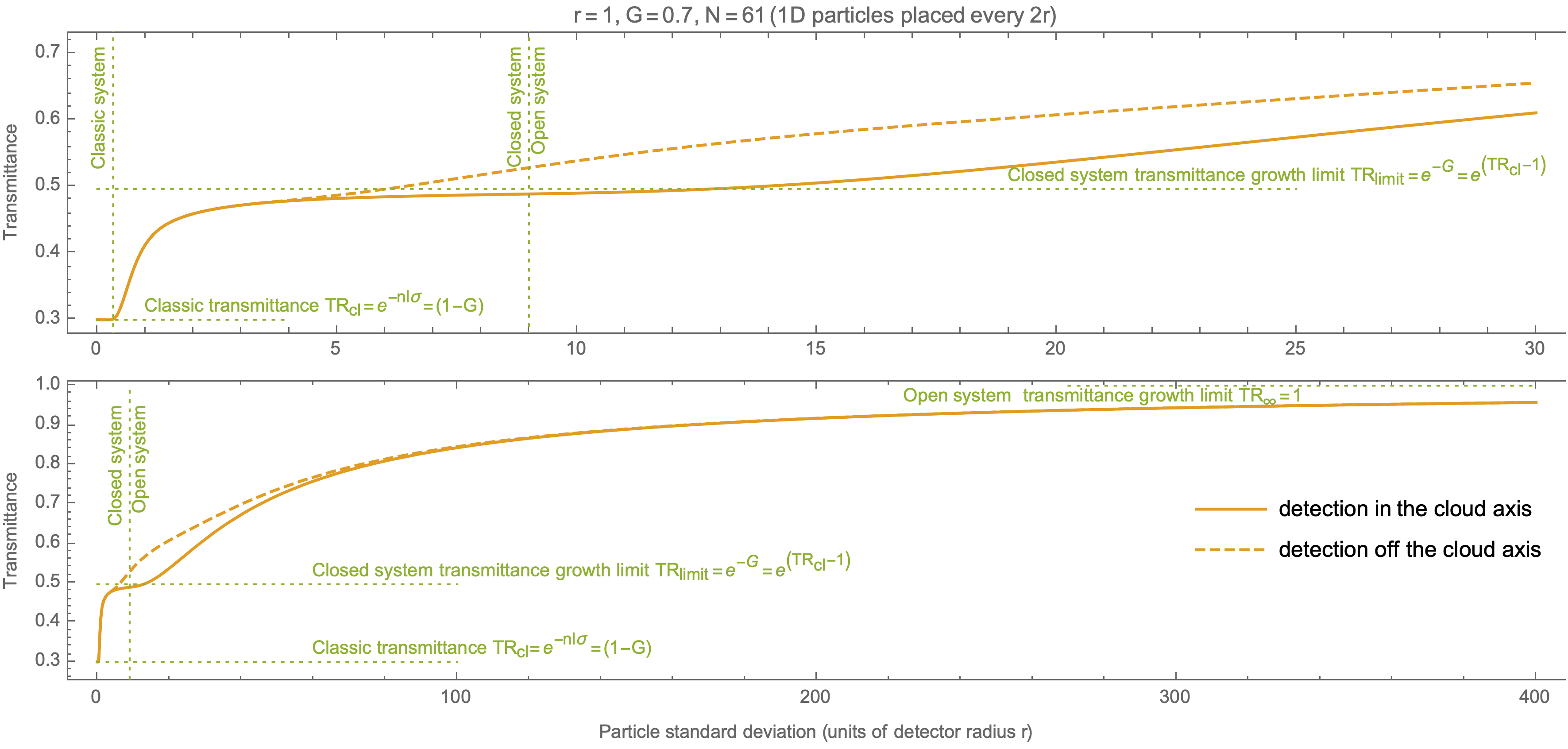}
\caption{The charts show the sample dependence of transmittance on the standard deviation for the measurement with a fixed size detector. The solid line is the measurement in the cloud axis, and the dashed line denotes some off-axis measurement. The length unit is equal to detector radius $r$. The detector width is 2 ($r=1$). A 1D cloud is $N=61$ particles in total, and they are evenly spaced every $2r$. The $G$ coefficient is set to 0.7 (after $TR_{cl}=30\%$). The particles have a 1D normal distribution where the standard deviation is denominated in detector radius units ($r$). The detector offset for off-axis detection is $20r$ from the cloud axis. The upper chart shows the magnification of the left part of the bottom diagram.} \label{fig:transmittancesChart}
\end{figure*}

Fig.~(\ref{fig:transmittancesChart}) shows the dependence of transmittance on the standard deviation for the measurement with a fixed size detector. The following section describes the chart details.

\subsection{Dense or inhomogeneous clouds}
If there are many more than one particle per detector (like in any real-world setup), then we repeat the above reasoning many times in the following way. We divide the gas cloud into enough parts so that each of them contains statistically only one particle per ''detector area``. We calculate the (partial) transmittance for each of these parts independently. From the property of independence of the probability of absorption by individual gas particles, we calculate the product of partial transmittances, obtaining the total transmittance of the entire cloud. 

The same approach works for analyzing inhomogeneous gas clouds. One should divide such a cloud into homogenous parts, calculate separately (partial) transmittances, and take their product to get total transmittance.

Alternatively, we can do the trick of adjusting the constant $G$. We can set it equal to $1-TR_{cl}$, where $TR_{cl}$ is the classic transmittance of the cloud. Then we take a set of ''artificial`` particles distributed exactly every $2r$ as requested above. Such an artificial particle represents all real particles present in the visibility tunnel of a given chunk. Remember, however, the spread (e.g., the standard deviation) of this artificial particle is the same as any single cloud particle. I.e., we don't sum individual particle masses to calculate spread speed. The latter method is very efficient for numerical computations.

\subsection{3D cloud}
For a three-dimensional gas cloud, it first has to be cast onto the plane of the detector. For such a 2D model, we demand to distribute particles evenly: one particle per detector area. The straightforward way to analyze 2D is to consider normal distribution and a square detector with a side equal to $2r$. For such a system: i) an analytical solution is available, see (11) and (18) in \cite{Ratajczak2020b} and ii) the square shape of the detector allows to cover the entire plane with adjacent detectors. Then, we can perform the same periodic reasoning as given above for the 1D model. 

An arbitrary shape of the detector makes reasoning more challenging and changes quantitative equations. Yet, it is still possible as it only requires a finite area of the detector. However, qualitatively the presented principle of the transmittance dependence on the detector area holds.

\section{Scope of applicability}
 
\subsection{Classic system}
The ideal gas is the classical limit of the model. Particles of such a gas have a negligible spread, and a detector is of macroscopic size ($r \gg stdev$) \cite{A.D.McNaught1997}. The upper chart of Fig.~(\ref{fig:transmittancesChart}) shows this limit on the left, close to $stdev = 0$. On the bottom chart of this figure, the area of the classical system applicability is practically invisible. 

\subsection{Open system}
An open system is a configuration where particle spread can reach a very high value, causing probability to leak out far from the cloud. We know, however, that in real physical systems, the maximum spread is limited. At least two factors limit $stdev$ growth: i) the age of the Universe and ii) the cloud environment causing particle decoherence (collapse of wave functions). It seems, however, that there may be conditions in outer space where decoherence is negligible (darkness and high vacuum), so the age of the Universe is the only upper limit. Particles of atomic size can experience a very significant spread there. Spread (e.g., measured with $stdev$) may be many orders of magnitude greater than a detector size. As a result, a considerable transmittance increase may happen there.
The upper limit of transmittance growth in the open system is 100\% simply because probability (mass) leaks out of the system.

\subsection{Closed system}
When the gas cloud is vast and the particles' spread is low, the probability doesn't leak out beyond the cloud's outline. Let us call it a closed system because it is, in fact, similar to closed systems with mass conserved.

In such a setup, the probability of distant particles flows into the measurement region. E.g., the gas may be closed in a sufficiently large chamber, i.e., its diameter $D_{chmb} \gg stdev$. Our experimental setup \cite{Ratajczak2021}, where $D_{ chmb} \sim \SI{25}{\centi\metre}$, $stdev \sim \SI{14}{\micro\metre}$ and $r \sim \SI{25}{\micro\metre}$ ($stdev \approx 0.56r$) is such an example. A closed system can also mean an open system (e.g., in outer space) but with a cloud diameter much larger than the spread: $D_{cloud} \gg stdev$. The cloud diameter is the diameter of the volume non-spread cloud occupies in a classical situation when $stdev \to 0$. An imaginary infinite system, i.e., particles placed infinitely far in all directions (perpendicular to the measurement axis), should also be considered a closed system.

The classic ideal gas is a special case of a closed system.

The green dashed line marked ''Close system/Open system`` on Fig.~(\ref{fig:transmittancesChart}) indicates the approximate boundary between close and open systems.

\subsubsection{Transmittance growth limit in a closed system}
We found there is a limit of transmittance growth in a closed system. Generally, the increase is possible thanks to Eq.~(\ref{TRUnfold})  product components factorization getting more uniform with increasing spread.  All components tend to 1: $(1-G/K) \to 1^{(-)}$ because the probability distribution spreading does not change the area under the curve. $K$ is a number of chunks that hold $P>0$. $K$ grows with the growth spread ($stdev$) so $G/K \to 0^{(+)}$, because $G=const$. For large $K$ we can rewrite Eq.~(\ref{TRUnfold}) considering only chunks with $P>0$ in the following way:
\begin{equation}
TR_{closed}=(1-G/K)^K
~. 
\end{equation}
We find there is upper limit of the last equation:
\begin{equation}
TR_{limit}=\lim_{K \to \infty}(1-G/K)^K=e^{-G}=e^{(TR_{cl}-1)}
~,
\end{equation}
because probability distribution gets divided into more and more ($K$) intervals. 

Reaching this limit is visible in Fig.~(\ref{fig:transmittancesChart}) as a curve flattening in the middle of the chart. The dashed green line labeled $TR_{limit}$ marks this limit. 

We conclude with an example. An almost entirely opaque (classically) cloud with $TR_{cl}\,=\,0$, (i.e., $G\,=\,1$) increases its transmittance (as a result of spontaneous spread but without mass escaping beyond the original cloud outline) up to a maximum of $TR_{limit}=e^{-1} \approx 36.8\%$.

\subsubsection{Measurement axis}
The boundary between the closed and the open system is not clearly specified, especially when the cloud is in an unlimited (deep) space. Carrying out the transmittance measurement in such a closed system closer to one of the cloud edges (instead of centrally coaxial) affects results. Less probability flows into the visibility tunnel from either direction because there are fewer particles. Therefore, the measured transmittance may be higher (for a specific particle spread) than when measured centrally through the cloud. 

The dashed curve Fig.~(\ref{fig:transmittancesChart}) shows such an off-axis measurement sample. It coincides with the solid line for the classic case (left-hand side). It is good. We don't expect any deviations of this type for ideal gas. Also, both lines overlap on the right-hand side because the open system doesn't have any specific axis. Only in the middle part of the chart, the dashed line is always above the solid line. It means that the transmittance measured closer to the edge of the cloud gets higher sooner (as $stdev$ grows).
This phenomenon may affect transmittance measurements of large deep space gas clouds.

\section{Distinguishability of QM interpretations}

The above analysis allows, among others, to experimentally distinguish some interpretations of quantum mechanics. In particular, the pilot-wave interpretation \cite{Bohm1952} assumes the existence of some localized objects that are only ''controlled`` by non-local functions. If it were so, if there were some kind of ''balls`` in the system with a certain probability (aka cross-section) of absorbing-or-not photons, then the dependence of transmittance on the spread would look different. In particular, the factorization of probabilities distributions would not affect the transmittance of a closed system. It would not lead to an increase in transmittance and reaching the $TR_{limit}=e^{-G}$ limit. Instead, in a closed system, the transmittance would apply according to the classical law of transmittance (Beer-Lambert law). A probability leak could be a cause of the only possible change in transmittance. However, it happens only in an open system.

\begin{figure*}
\includegraphics[width=\textwidth]{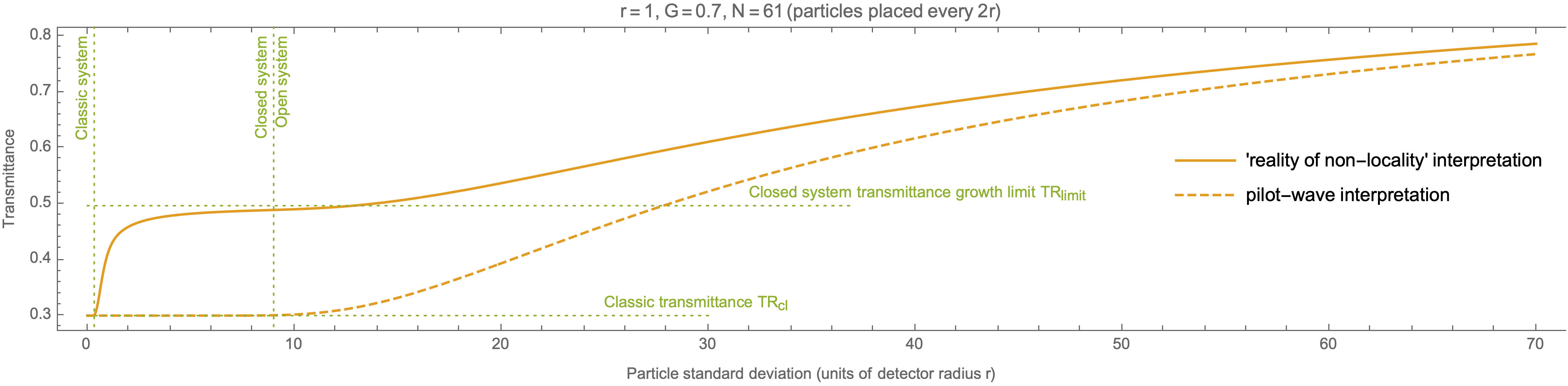}
\caption{Different QM interpretations lead to different smeared gas transmittance model predictions. The solid line indicates the transmittance predicted with the assumption of ''non-locality reality``. The dashed line marks the transmittance assuming there are some small, ball-like objects absorbing-or-not photons, i.e., according to the pilot-wave interpretation. The pilot-wave interpretation does not reveal any difference from classic transmittance for systems with mass conserved (see ''Closed system`` range).
} \label{fig:interpretationsChart}
\end{figure*}

Fig.~(\ref{fig:interpretationsChart}) shows this difference. The solid line indicates the transmittance predicted with the assumption of ''non-locality reality``. It doesn't expect any finite size, ball-like objects to exist. The dashed line marks the transmittance assuming some small, ball-like objects absorbing-or-not photons and that non-local wave functions just guide them. The dashed line shows the transmittance according to the pilot-wave interpretation. One can see the distance between the two graphs, which indicates the possibility of conducting experiments comparing the transmittances (such as \cite{Ratajczak2021}) and thus differentiating the interpretations.

\section{Summary}
This analysis extends the transmittance analysis of ultra-diluted gases presented in \cite{Ratajczak2020b}. It shows the non-obvious influence of particles' quantum spread and detector area for optical transmittance measurement. It predicts that the optical transmittance of the smeared gas cloud may change (raise) even when the system's mass is conserved, like in some laboratory experiments. We found the limit of such growth. The presented mathematical analysis does not depend on any specific form of the gas-particle wave function. The paper also presents a brief analysis of the possibilities of distinguishing between some interpretations of quantum mechanics. The model is falsifiable. We propose possible experiments in \cite{Ratajczak2020b} and report promising results of one of them in \cite{Ratajczak2021}.

This model may help better understand phenomena occurring in deep space. A dark vacuum has natural conditions for the spontaneous formation of smeared gas out of ideal gas. Diluted gas is one of the most abundant forms of matter in the Universe. Observation of its transmittance, such as spectroscopy, is one of the essential tools for studying its properties: composition, density, changes, etc. This theory may be of some importance for the correct interpretation of astronomical observations and astrophysical models. In addition, the demonstrated tendency for a spontaneous increase in transmittance may be a part of the answer to the problem of the missing visible mass in the Universe, the so-called dark matter.

\section{Data availability}
All data generated or analysed during this study are included in this published article (and its supplementary information files).

\bibliography{../library.bib}

\end{document}